\documentclass[11pt]{article}

\usepackage[margin=1in]{geometry}
\usepackage[T1]{fontenc}
\usepackage[utf8]{inputenc}
\usepackage{lmodern}
\usepackage{microtype}
\usepackage{amsmath,amssymb,amsfonts}
\usepackage{booktabs}
\usepackage{graphicx}
\usepackage{caption}
\usepackage{subcaption}
\usepackage{siunitx}
\usepackage{tabularx}
\usepackage{multirow}
\usepackage{enumitem}
\usepackage{hyperref}
\usepackage[numbers,sort&compress]{natbib}
\usepackage{tikz}
\usetikzlibrary{arrows.meta,positioning,fit,calc}
\hypersetup{
  colorlinks=true,
  linkcolor=blue,
  citecolor=blue,
  urlcolor=blue
}

\title{Topological Sensitivity in Connectome-Constrained Neural Networks}
\author{Nalin Dhiman\\School of Computing and Electrical Engineering\\Indian Institute of Technology, Mandi, India\\\texttt{d24008@students.iitmandi.ac.in}}
\date{}

\begin{document}
\maketitle

\begin{abstract}
Connectome-constrained neural networks are often evaluated against sparse random controls and then interpreted as evidence that biological graph topology improves learning efficiency. We revisit that claim in a controlled flyvis-based study using a Drosophila connectome, a naive self-loop-matched random graph, and a degree-preserving rewired null. Under weak controls, in which both models were recovered from a connectome-trained checkpoint and the null matched only global graph counts, the connectome appeared substantially better in early loss, mean activity, and runtime. That picture changed under stricter controls. Training both graphs from a shared random initialization removed the early loss advantage, and replacing the naive null by a degree-preserving null removed the apparent activity advantage. A five-sample degree-preserving ensemble and a pre-training activity-scale diagnostic further strengthened this revised interpretation. We also report a descriptive mechanism analysis of the earlier weak-control comparison, but we treat it as behavioral characterization rather than proof of causal superiority. We show that previously reported topology advantages in connectome-constrained neural networks can arise from initialization and null-model confounds, and largely disappear under fair from-scratch initialization and degree-preserving controls.
\end{abstract}

\section{Introduction}

Graph topology is a natural source of inductive bias in sparse neural systems. In machine learning, structured connectivity can influence optimization, representational bias, and computational cost even when parameter count is held fixed \citep{battaglia2018relational,bronstein2021geometric,frankle2019lottery,mocanu2018scalable,xie2019randomly}. In neuroscience, connectomes provide concrete wiring diagrams rather than abstract graph models, offering a direct way to test whether biologically derived topology contributes useful computational structure \citep{scheffer2020adult,borst2010fly,sanes2010design,bullmore2009complex,bassett2017network}.

This question is particularly compelling in the fly visual system, where direction-selective pathways and their anatomical organization are well characterized \citep{takemura2013motion,maisak2013directional,joesch2010onoff,clark2011defining,eichner2011motiondetector,behnia2014processing,silies2013modular,borst2015common}. A natural hypothesis is that a connectome-constrained network derived from this circuitry may learn more efficiently than a sparse random control.

However, attributing differences to topology alone is nontrivial. Sparse graph comparisons are easily confounded by initialization, degree sequence, and parameter mapping. For example, initializing a control graph from a checkpoint adapted to a specific topology or using a null model that fails to preserve degree structure can produce apparent “topology effects” that do not reflect the graph itself \citep{newman2001randomgraphs,maslov2002specificity,milo2002network,rubinov2010complex}. Similar sensitivities are well documented in sparse-network training, where initialization and connectivity interact strongly with early optimization dynamics \citep{bellec2018deeprewiring,evci2020rigging,mostafa2019dynamic,hoefler2021sparsity}.

This study addresses that issue through a structured sequence of controls, which we refer to as a \emph{control ladder}. We begin from an empirical observation that appears strong: under a checkpoint-based comparison against a naive random graph, the connectome model shows lower early loss, lower mean activity, and faster runtime. We then progressively remove two key confounds. First, we eliminate checkpoint bias by training both graphs from a shared random initialization. Second, we strengthen the null model by replacing the naive random graph with a degree-preserving rewired graph that matches directed in-degree and out-degree sequences in addition to global graph statistics.

Our results revise the initial interpretation. The apparent connectome advantage does not persist under stricter controls: training from a shared random initialization removes the early loss gap, and using a degree-preserving null removes the activity difference. These findings are consistent across a multi-sample degree-preserving ensemble and are not explained by differences in initial activity scale. Together, they show that the originally observed advantage can be accounted for by initialization and null-model design, rather than by topology alone.

\begin{figure}[t]
  \centering
  \includegraphics[width=\linewidth]{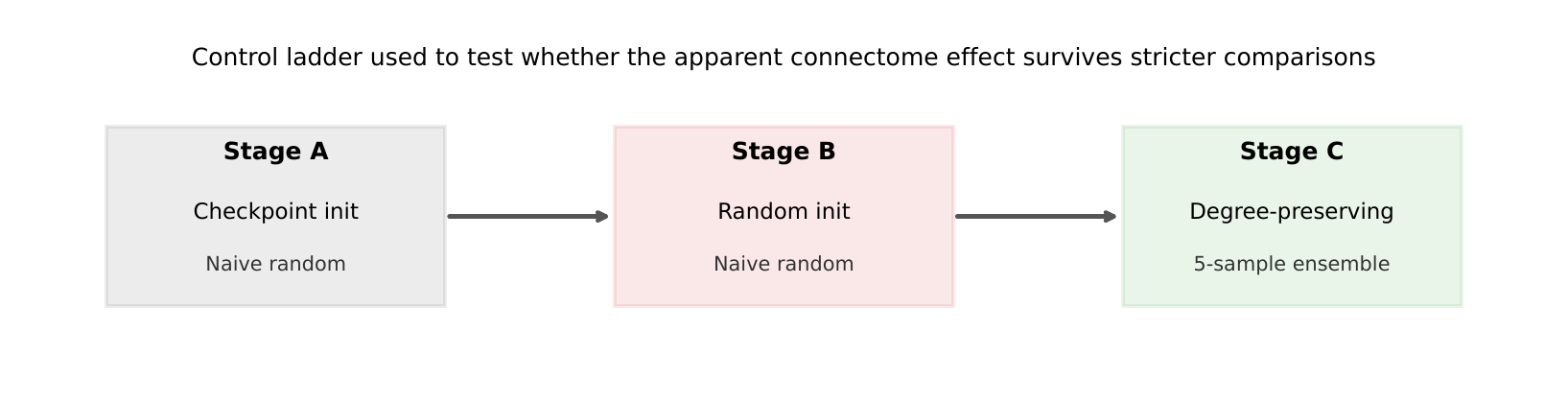}
  \caption{Control ladder used in the revision study. The original observation compared a connectome graph to a self-loop-matched random graph after checkpoint recovery. The corrected analysis then removed checkpoint initialization and strengthened the null model by preserving the directed degree sequence. The substantive scientific conclusion changes across these control levels.}
  \label{fig:control-diagram}
\end{figure}

Figure~\ref{fig:control-diagram} summarizes the control ladder that drives the analysis. The remainder of the paper formalizes this comparison, presents the corrected results, and revisits an earlier mechanism analysis in appropriately limited terms.

\section{Related Work}

Our study sits at the intersection of connectomics, sparse neural network design, and null-model methodology.

\paragraph{Connectomics and fly motion circuits.}
The fly visual system has become a standard model for circuit-level analyses of motion computation \citep{borst2010fly,sanes2010design}. Anatomical and functional work has mapped many of the pathways that feed elementary motion detection and downstream direction-selective populations \citep{joesch2010onoff,clark2011defining,eichner2011motiondetector,takemura2013motion,maisak2013directional,behnia2014processing,silies2013modular,borst2015common}. Large-scale Drosophila connectomics has made it possible to treat the wiring diagram itself as a modeling object \citep{scheffer2020adult}. This motivates connectome-constrained modeling, but does not by itself establish that empirical wiring outperforms appropriate nulls.

\paragraph{Biologically inspired and connectome-based neural modeling.}
There is longstanding interest in using neuroscience to inform machine learning architectures \citep{marblestone2016toward,hassabis2017neuroscience,richards2019deep,yamins2016using,zador2019critique}. Some work attempts to import circuit motifs directly, whereas other work uses biological connectivity as a prior or a constraint \citep{billeh2020systematic,lechner2020neural}. These efforts are valuable, but they also highlight a recurring issue: biologically grounded structure is often changed together with initialization, parameterization, or task setup, making it difficult to attribute performance differences to topology alone.

\paragraph{Sparse networks, structured connectivity, and optimization.}
The sparse-network literature shows that connectivity structure can matter, but also that training behavior depends on how sparsity is introduced and maintained \citep{frankle2019lottery,mocanu2018scalable,bellec2018deeprewiring,mostafa2019dynamic,evci2020rigging,hoefler2021sparsity,gale2019state}. Randomly wired networks can themselves exhibit meaningful inductive biases depending on the graph family used \citep{xie2019randomly}. These results motivate controlled graph comparisons, but they also warn against attributing early training differences to topology without testing alternative nulls.

\paragraph{Network science and null models.}
Graph comparisons are meaningful only relative to the null they employ \citep{watts1998collective,barabasi1999emergence,newman2003structure,bullmore2009complex,bullmore2012economy,bassett2017network}. In directed biological networks, preserving only node and edge counts is usually insufficient; degree sequence, self-loops, and local motifs can all materially change conclusions \citep{newman2001randomgraphs,milo2002network,maslov2002specificity,rubinov2010complex,sporns2004small,kaiser2006nonoptimal}. Our revision is directly informed by this literature: the degree-preserving rewired graph is intended as a stricter null than the original naive random graph.

\paragraph{Efficient computation and activity costs.}
The connectome comparison in this study was initially motivated by activation efficiency, which relates to efficient coding and metabolic-cost perspectives in neuroscience \citep{barlow1961possible,olshausen1996emergence,vinje2000sparse,simoncelli2001natural,attwell2001energy,lennie2003cost,laughlin2001energy,carandini2012normalization,deneve2016efficient,sterling2015principles}. However, lower activity by itself does not establish a causal or principled advantage of a given topology. In this work, activity is therefore treated as an empirical quantity that must be interpreted alongside appropriate controls, rather than as evidence of optimality.
We show that the apparent advantage observed under weak controls changes substantially once two specific confounds—checkpoint initialization and a weak random null—are removed. This isolates the role of initialization and null-model design in shaping conclusions about topology.

\usetikzlibrary{positioning, fit, arrows.meta, calc}

\begin{figure}[t]
\centering
\resizebox{\textwidth}{!}{%
\begin{tikzpicture}[
    font=\small,
    >=Latex,
    box/.style={draw, rounded corners, thick, align=center,
                minimum height=12mm, minimum width=32mm},
    smallbox/.style={draw, rounded corners, thick, align=center,
                     minimum height=9mm, minimum width=44mm},
    note/.style={align=left, text width=55mm, font=\small},
    arrow/.style={-Latex, thick},
    line/.style={thick}
]

%% ── Input ───────────────────────────────────────────────────────────────
\node[box, fill=gray!8] (stim)
    {MovingEdge\\stimulus sequence\\$x^{(t)}$};

%% ── Mask nodes ──────────────────────────────────────────────────────────
\node[smallbox, fill=blue!8,  right=30mm of stim, yshift=16mm] (conn)
    {Connectome mask $G_{\mathrm{conn}}$};
\node[smallbox, fill=red!8,   below=6mm of conn]               (rand)
    {Naive random mask $G_{\mathrm{rand}}$};
\node[smallbox, fill=green!8, below=6mm of rand]               (deg)
    {Degree-preserving mask $G_{\mathrm{degpres}}$};

%% Dashed enclosure
\node[draw, dashed, rounded corners,
      fit=(conn)(rand)(deg), inner sep=6mm,
      label={[font=\small\color{blue!60!black}]above:Graph constraint / null model}]
      (maskfit) {};

%% ── Core: anchored RIGHT of maskfit (never of rand) ────────────────────
\node[box, fill=yellow!15,
      right=35mm of maskfit,
      minimum width=60mm, minimum height=22mm] (core)
    {Graph-masked recurrent core\\[3mm]
     $\mathbf{h}^{(t+1)}=\sigma\!\left(W_G\mathbf{h}^{(t)}+\mathbf{x}^{(t)}\right)$};

%% ── Downstream nodes ────────────────────────────────────────────────────
\node[box, fill=gray!8,    right=28mm of core, minimum width=40mm] (pool)
    {Central-cell selection\\and temporal pooling};
\node[box, fill=orange!12, right=20mm of pool, minimum width=34mm] (dec)
    {Linear decoder\\$\hat{\mathbf{y}}=D\,\phi(\mathbf{h})$};
\node[box, fill=gray!8,    right=20mm of dec,  minimum width=34mm] (out)
    {2D direction target\\$(\cos\theta,\sin\theta)$};

%% ── stim → core: goes BELOW the mask group ──────────────────────────────
%% Down from stim, east under maskfit, up into core.south
\draw[arrow] (stim.south)
    |- ([yshift=-12mm]deg.south)     %% drops below the lowest mask node
    -| (core.south);                 %% comes up into the core from below

%% ── Downstream pipeline ─────────────────────────────────────────────────
\draw[arrow] (core) --
    node[above, font=\scriptsize, midway]{central-cell features}
    (pool);
\draw[arrow] (pool) -- (dec);
\draw[arrow] (dec)  -- (out);

%% ── Mask fan-in using COLLECTOR SPINE technique ─────────────────────────
%% Step 1: horizontal stubs from each mask node to a vertical spine
%%         The spine sits at maskfit.east + 10mm (clearly outside the box)
\coordinate (spineX) at ($(maskfit.east)+(10mm,0)$);

\draw[line] (conn.east) -- (conn.east -| spineX);
\draw[line] (rand.east) -- (rand.east -| spineX);
\draw[line] (deg.east)  -- (deg.east  -| spineX);

%% Step 2: vertical spine connecting top stub to bottom stub
\draw[line] (conn.east -| spineX) -- (deg.east -| spineX);

%% Step 3: single arrow from spine midpoint → core.west
%%         Midpoint = vertical centre of rand (middle mask node)
\draw[arrow] (rand.east -| spineX) -- (core.west);

%% ── Notes ───────────────────────────────────────────────────────────────
%% ── Notes: side by side, well below the diagram ────────────────────────
%% Anchor both notes to a point below the FULL diagram (below deg/maskfit)
%% and place them side by side, not stacked.

\node[note, text width=60mm,
      below=28mm of deg, anchor=north west,
      xshift=-10mm] (params)
{%
    \textbf{Parameterization:}\\[3pt]
    734 trainable parameters, 2959 fixed parameters.\\[4pt]
    Rewiring changes graph topology while preserving
    the shared parameterization scheme.%
};

\end{tikzpicture}%
}
\caption{\textbf{Architecture used in the study.}
A MovingEdge stimulus drives a graph-constrained recurrent \texttt{flyvis}
network whose connectivity mask is given by either the empirical connectome,
a naive random graph, or a degree-preserving random graph.
A linear decoder reads pooled central-cell activity and predicts the 2D
motion direction target. The key comparison in the paper is not between
different decoders or optimizers, but between different graph masks under
matched nodes, edges, self-loops, and parameterization.}
\label{fig:architecture}
\end{figure}
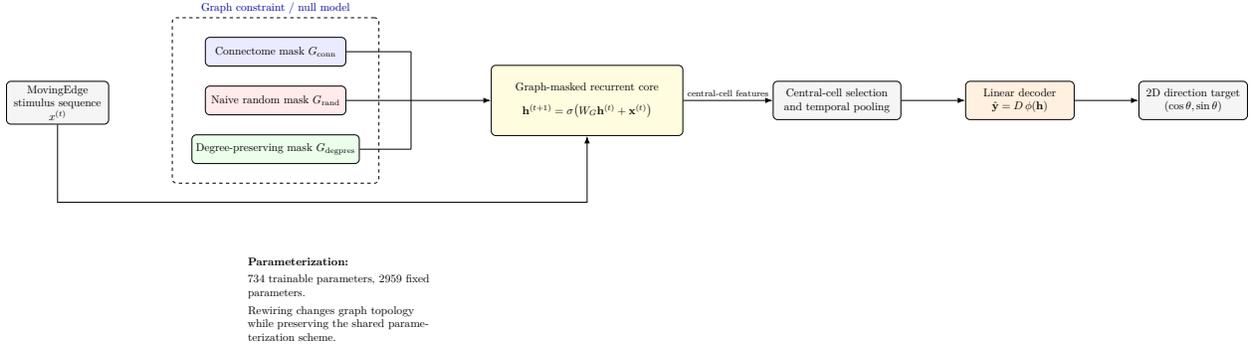
\section{Problem Formulation}

Let $G = (V, E)$ be a directed graph with node set $V$ and edge set $E$. We compare three graph families:
\begin{align}
  G_{\mathrm{conn}}, \qquad
  G_{\mathrm{rand}}, \qquad
  G_{\mathrm{degpres}},
\end{align}
where $G_{\mathrm{conn}}$ is the empirical connectome, $G_{\mathrm{rand}}$ is a self-loop-matched random graph, and $G_{\mathrm{degpres}}$ is a degree-preserving rewired graph.

The comparison is constrained so that
\begin{align}
  |V_{\mathrm{conn}}| = |V_{\mathrm{rand}}| = |V_{\mathrm{degpres}}|, \\
  |E_{\mathrm{conn}}| = |E_{\mathrm{rand}}| = |E_{\mathrm{degpres}}|, \\
  \ell_{\mathrm{conn}} = \ell_{\mathrm{rand}} = \ell_{\mathrm{degpres}},
\end{align}
where $\ell$ denotes the number of self-loops. The flyvis network implementation also keeps the same network parameterization under rewiring, with 734 trainable network parameters and 2959 fixed network parameters across all graph conditions. A separate linear decoder is trained identically in every condition.

We write the recurrent network state as $h^{(\tau)} \in \mathbb{R}^{|V|}$ for internal time index $\tau$ within a stimulus sequence. Abstracting the implementation, the state update can be written as
\begin{align}
  h^{(\tau+1)} = \sigma\!\left(W_G h^{(\tau)} + u\!\left(x^{(\tau)}\right)\right),
\end{align}
where $x^{(\tau)}$ is the external stimulus, $\sigma(\cdot)$ is the model nonlinearity, and $W_G$ respects the graph adjacency induced by $G$. More explicitly, if $M_G \in \{0,1\}^{|V|\times|V|}$ is the adjacency mask, then
\begin{align}
  W_G = M_G \odot \widetilde{W}(\theta,\phi),
\end{align}
where $\theta$ are trainable network parameters and $\phi$ are fixed network parameters inherited from the flyvis parameter tables.

For a batch of stimuli, the decoder receives pooled central-cell activity and predicts a two-dimensional motion target. Let $\hat{y}(\theta_t)$ denote the decoder output after training step $t$ and let $y$ denote the target direction vector. The task loss is mean squared error,
\begin{align}
  \mathcal{L}(\theta_t) = \mathbb{E}\big[\|\hat{y}(\theta_t) - y\|_2^2\big].
\end{align}

Training follows a matched-step protocol:
\begin{align}
  \theta_{t+1} = \theta_t - \eta \nabla \mathcal{L}(\theta_t),
\end{align}
with identical optimizer type, learning rate, task batch, and number of updates across compared models. We evaluate at fixed horizons $T \in \{5,10\}$ rather than at matched wall-clock time.

The activity metric is mean absolute activation,
\begin{align}
  A(\theta_t) = \mathbb{E}_{b,\tau,i}\left[\,|h_{b,\tau,i}|\,\right].
\end{align}
For saved post-training activity tensors we also define per-node activity
\begin{align}
  a_i = \mathbb{E}_{b,\tau}\left[\,|h_{b,\tau,i}|\,\right],
\end{align}
the Gini coefficient
\begin{align}
  \mathrm{Gini}(a) =
  \frac{\sum_{i=1}^{n}\sum_{j=1}^{n}|a_i-a_j|}{2n\sum_{i=1}^{n} a_i},
\end{align}
and an edge-usage proxy
\begin{align}
  u_{ij} = |w_{ij}| \; \mathbb{E}_{b,\tau}\left[\,|h_{b,\tau,j}|\,\right].
\end{align}

The central question is whether an apparent connectome advantage persists under progressively stricter controls on initialization and null-model design.
\section{Methods}

\subsection{Graph Construction}

All experiments are conducted on a fixed flyvis network scaffold with a Drosophila-derived node set. The empirical connectome graph $G_{\mathrm{conn}}$ contains 45{,}669 nodes, 1{,}513{,}231 directed edges, and 12{,}380 self-loops. Across all graph conditions, the recurrent network exposes 734 trainable parameters and 2959 fixed parameters, and an identical linear decoder is trained in every case.

We consider two control graph families.

The \emph{naive random} control $G_{\mathrm{rand}}$ is constructed by replacing the edge incidence list with uniformly sampled self-loop-matched pairs. Self-loops are preserved exactly, and the remaining edges are sampled without replacement from all ordered node pairs. This matches node count, edge count, density, self-loop count, and parameter count, but does not preserve directed degree sequence.

The \emph{degree-preserving} control $G_{\mathrm{degpres}}$ is constructed by directed double-edge swaps applied to the non-loop edges of the connectome while holding the self-loops fixed. Each accepted swap replaces $(a,b)$ and $(c,d)$ with $(a,d)$ and $(c,b)$ under the constraints that no self-loops or duplicate edges are introduced and that both in-degree and out-degree sequences are exactly preserved. To assess robustness, we generate an ensemble of five independently rewired graphs. Each sample performs 250{,}000 accepted swaps (approximately $0.17$ swaps per edge), with an acceptance rate of $0.992$ and no duplicate edges. All graph-level quantities remain exactly matched (Table~\ref{tab:graph-matching}).

\begin{table}[t]
  \centering
  \caption{Graph-level matching constraints used throughout the study. The recurrent network parameterization is identical across all graph conditions.}
  \label{tab:graph-matching}
  \begin{tabular}{lrrrrr}
    \toprule
    Graph & Nodes & Edges & Self-loops & Trainable & Fixed \\
    \midrule
    Connectome & 45{,}669 & 1{,}513{,}231 & 12{,}380 & 734 & 2959 \\
    Naive random & 45{,}669 & 1{,}513{,}231 & 12{,}380 & 734 & 2959 \\
    Degree-preserving random & 45{,}669 & 1{,}513{,}231 & 12{,}380 & 734 & 2959 \\
    \bottomrule
  \end{tabular}
\end{table}

\subsection{Parameterization and Rewiring Semantics}

The flyvis model employs a parameter-sharing scheme defined by edge and node metadata tables. The recurrent network is parameterized by a small set of shared weights (734 trainable and 2959 fixed), which are reused across many edges. Rewiring modifies only the source-target incidence structure while leaving the metadata tables and parameter assignments unchanged. Consequently, all graph conditions share identical parameter counts, parameter values, and decoder construction, while differing only in how those parameters are arranged over the graph. This preserves comparability while maintaining consistency with the underlying implementation.

\subsection{Initialization Schemes}

We consider two initialization regimes.

\paragraph{Checkpoint-based initialization.}
In the original comparison, both the connectome and naive random graphs were initialized from a checkpoint trained on the connectome topology. This introduces a bias because parameters are already adapted to the connectome structure.

\paragraph{From-scratch initialization.}
In the corrected experiments, all graph conditions are initialized directly from the same topology-agnostic initialization procedure. No checkpoint recovery is used. Each experiment uses a shared random seed so that parameter initialization is aligned across graph types as closely as possible.

\subsection{Decoder and Optimization}

A linear decoder maps pooled central-cell activity to a two-dimensional motion direction target $(\cos\theta,\sin\theta)$. Optimization uses Adam with a learning rate of $10^{-3}$, identical across all graph conditions. No hyperparameter tuning is performed between conditions.

\subsection{Descriptive Mechanism Metrics}

We compute three families of descriptive diagnostics on saved post-training activity tensors:

\begin{enumerate}[leftmargin=1.5em]
  \item \textbf{Node activity concentration:} per-node mean absolute activity, total activity, Gini coefficient, and top-$k$ contribution fractions.
  \item \textbf{Edge usage proxy:} $u_{ij} = |w_{ij}|\,\mathbb{E}|h_j|$, with corresponding concentration metrics.
  \item \textbf{Temporal summaries:} total activity over time, variance across nodes, and mean absolute timestep-to-timestep change.
\end{enumerate}

These quantities are used to characterize observed dynamics. They are not interpreted as causal evidence of a topology advantage.
\section{Experimental Setup}

\subsection{Task and Canonical Batch}

All experiments use the canonical Stage 3 flyvis task: MovingEdge direction decoding. The training batch consists of 12 stimuli with 269 frames each and spatial input shape $(12, 269, 1, 721)$. Training is performed on the corresponding training split with speed 2.4, angles $\{0,60,120,180,240,300\}$, and intensities $\{0,1\}$. The study focuses on matched-step comparisons on this canonical batch to isolate early-learning behavior under controlled conditions.

\subsection{Matched-Step Protocol}

Models are compared after exactly 5 and 10 optimization steps. This short-horizon protocol isolates early optimization dynamics by matching update count rather than wall-clock time. All conditions use the same training batch, decoder, optimizer, and experimental seeds $\{0,1,2\}$.

\subsection{Control Ladder}

Experiments are organized as a sequence of progressively stricter controls.

\paragraph{Stage A: checkpoint-based comparison.}
The connectome is compared to a naive random graph under checkpoint initialization.

\paragraph{Stage B: initialization control.}
Both graphs are trained from a shared from-scratch initialization, removing checkpoint bias.

\paragraph{Stage C: degree-preserving control.}
The connectome is compared to a degree-preserving rewired graph under the same from-scratch initialization. Both a single instance and an ensemble of five independent rewired graphs are evaluated.

\subsection{Metrics}

We report three primary metrics:
\begin{enumerate}[leftmargin=1.5em]
  \item \textbf{Loss:} decoder mean squared error at matched update counts,
  \item \textbf{Activity:} mean absolute network activity over batch, time, and nodes,
  \item \textbf{Runtime:} elapsed wall-clock time to reach the matched step horizon.
\end{enumerate}

Results are summarized across seeds by mean $\pm$ standard deviation. For the degree-preserving ensemble, we additionally report bootstrap confidence intervals for control-minus-connectome differences.

\subsection{Additional Diagnostics}

Two supplementary diagnostics are included:

\begin{enumerate}[leftmargin=1.5em]
  \item a degree-preserving ensemble consisting of five independently rewired graphs evaluated at 5 steps, and
  \item a pre-training activity-scale comparison across connectome, naive random, and degree-preserving graphs under the same initialization.
\end{enumerate}

The initial activity levels across graph conditions are closely matched, and no additional calibration is applied.
\section{Results}

\subsection{Original Observation Under Weak Controls}

Under checkpoint-based initialization and a naive self-loop-matched random null, the connectome exhibits lower loss, lower activity, and faster runtime at both 5 and 10 matched steps. At 5 steps, mean loss is $0.514$ for the connectome and $0.698$ for the naive random graph, mean activity is $0.656$ versus $1.861$, and elapsed time is $252$\,s versus $309$\,s. At 10 steps, the same ordering persists: loss $0.499$ versus $0.557$, activity $0.740$ versus $1.379$, and elapsed time $546$\,s versus $645$\,s.

\begin{figure}[t]
  \centering
  \includegraphics[width=\linewidth]{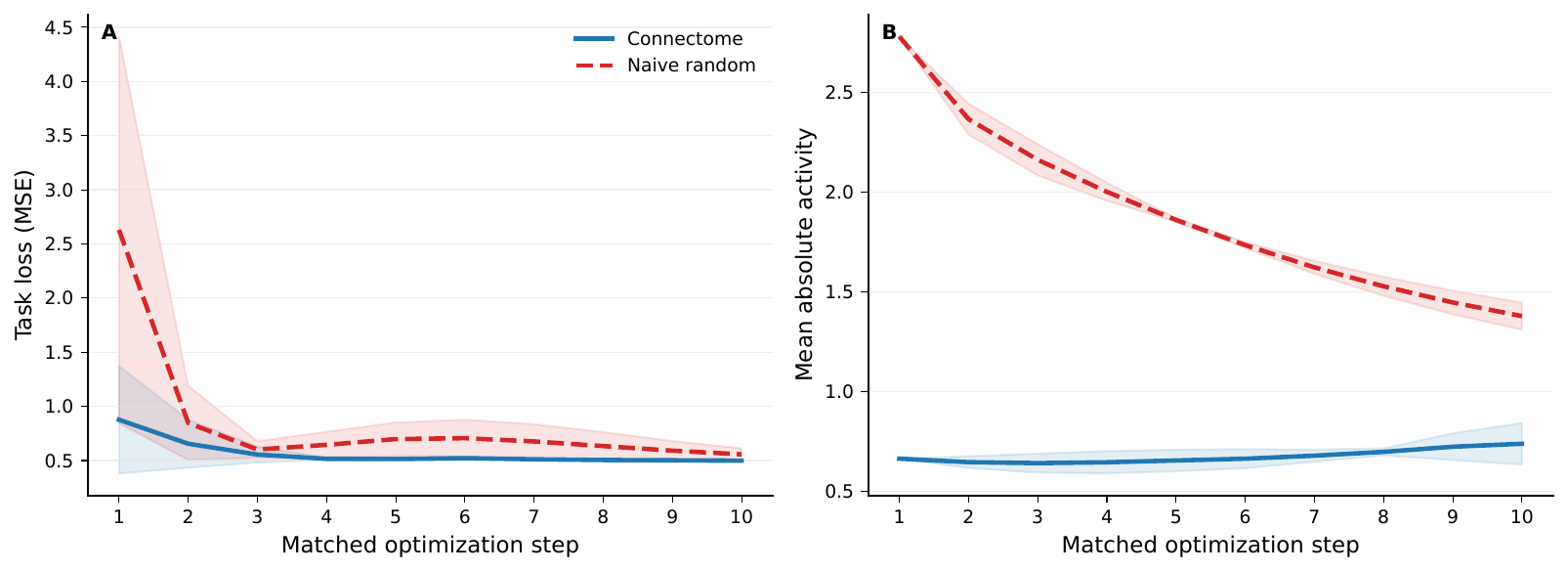}
  \caption{Matched-step training curves under checkpoint-based initialization and a naive random null. The connectome appears to outperform the control across all three metrics.}
  \label{fig:orig-curves}
\end{figure}

\begin{figure}[t]
  \centering
  \includegraphics[width=\linewidth]{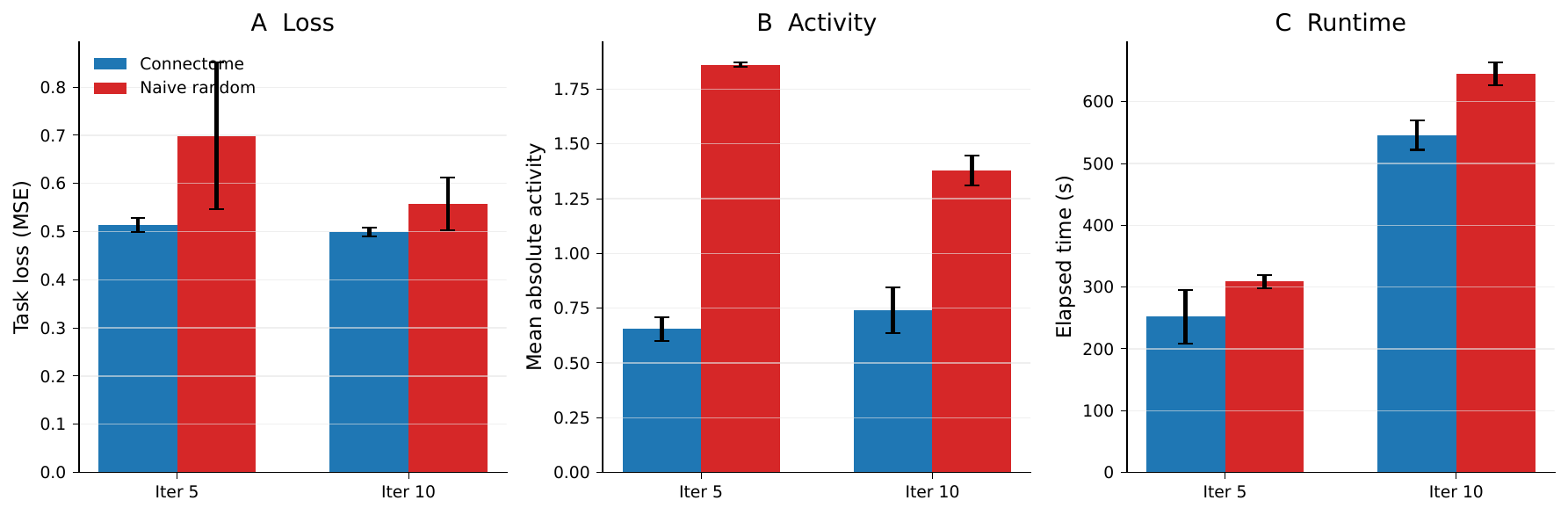}
  \caption{Matched-step summary at 5 and 10 steps under weak controls. All three metrics favor the connectome prior to applying stricter controls.}
  \label{fig:orig-summary}
\end{figure}

Figures~\ref{fig:orig-curves} and \ref{fig:orig-summary} summarize these observations. However, this comparison is confounded by checkpoint-based initialization and a null model that does not preserve directed degree sequence. The following controls isolate the effect of these factors.

\subsection{Initialization Control}

Training both graphs from a shared from-scratch initialization removes the loss advantage. At 5 steps, the loss difference between the naive random graph and the connectome is $-0.0020$, compared to $+0.1841$ under checkpoint initialization. At 10 steps, the loss difference remains $-0.0020$. 

Under this control, the connectome retains slightly lower activity and shorter runtime, but the primary loss advantage does not persist.

\subsection{Degree-Preserving Control}

Replacing the naive random null with a degree-preserving null removes the remaining activity advantage. At 5 steps, the loss difference is effectively zero ($+0.0003$), and the activity difference reverses sign ($-0.0106$), indicating slightly lower activity in the degree-preserving control. At 10 steps, both loss and activity differences remain close to zero, while the runtime difference is modest ($+3.98$\,s).

\subsection{Ensemble Robustness}

To assess robustness, we evaluate an ensemble of five independently rewired degree-preserving graphs across three seeds. Across all 15 sample-seed combinations, mean loss at 5 steps is $0.5155 \pm 0.0067$ for the connectome and $0.5172 \pm 0.0061$ for the degree-preserving control. Mean activity is $0.5453 \pm 0.0185$ for the connectome and $0.5346 \pm 0.0147$ for the control, while elapsed time is $122.75 \pm 2.17$\,s versus $133.47 \pm 7.05$\,s. 

The corresponding mean deltas are $+0.0017$ for loss, $-0.0108$ for activity, and $+10.72$\,s for elapsed time. These results are consistent across samples and do not recover the original advantage.

\begin{figure}[t]
  \centering
  \includegraphics[width=\linewidth]{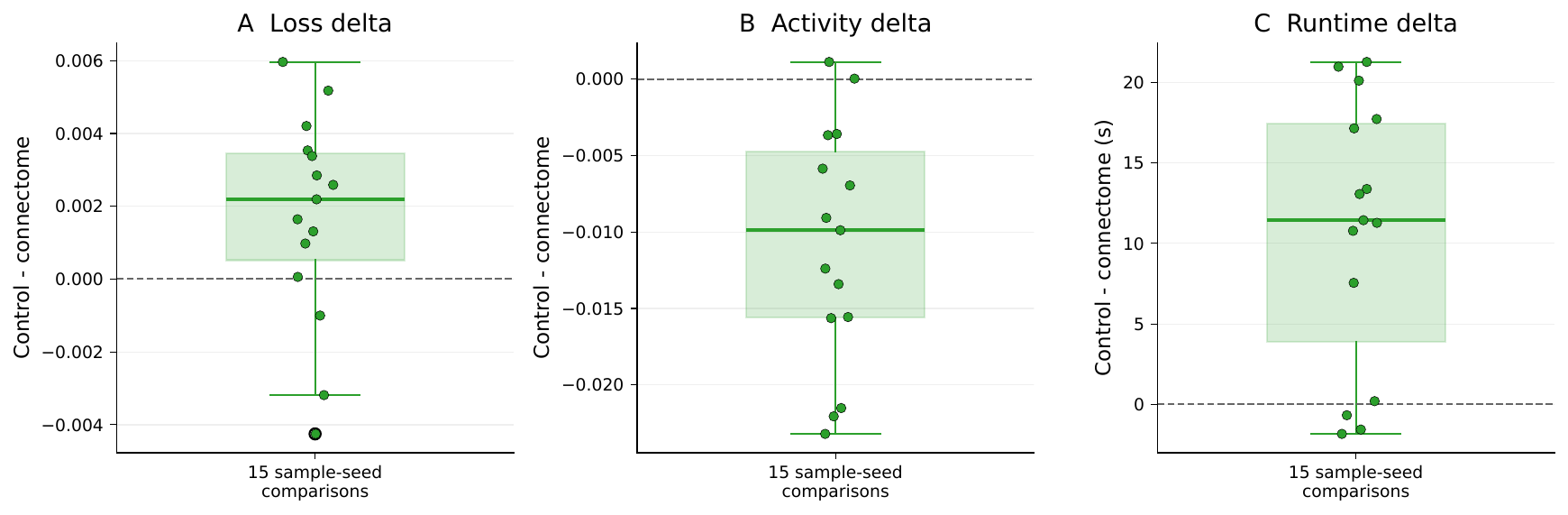}
  \caption{Degree-preserving ensemble variability at 5 matched steps. Each point represents one sample-seed comparison. Loss differences remain near zero, activity differences are slightly negative, and runtime differences are modest.}
  \label{fig:ensemble-variability}
\end{figure}

\subsection{Initial Activity Diagnostic}

To test whether residual differences reflect trivial scale mismatches, we compare initial activity and gradient norms prior to training. Under the same from-scratch initialization, mean absolute activity is $0.5682 \pm 0.0059$ for the connectome and $0.5774 \pm 0.0062$ for the degree-preserving control. The corresponding gradient norms are $9.75 \pm 4.21$ and $8.96 \pm 3.96$. These values indicate closely matched initial dynamical regimes.

\begin{table}[t]
  \centering
  \caption{Control progression across experiments. Deltas are control minus connectome.}
  \label{tab:control-progression}
  \begin{tabular}{lrrrr}
    \toprule
    Comparison & Steps & $\Delta$ loss & $\Delta$ activity & $\Delta$ elapsed (s) \\
    \midrule
    Original checkpoint + naive random & 5 & 0.1841 & 1.2056 & 57.00 \\
    Original checkpoint + naive random & 10 & 0.0583 & 0.6397 & 99.11 \\
    Random init + naive random & 5 & -0.0020 & 0.0323 & 57.60 \\
    Random init + naive random & 10 & -0.0020 & 0.0202 & 369.59 \\
    Random init + degree-preserving & 5 & 0.0003 & -0.0106 & 8.38 \\
    Random init + degree-preserving & 10 & -0.0018 & -0.0087 & 3.98 \\
    Degree-preserving ensemble & 5 & 0.0017 & -0.0108 & 10.72 \\
    \bottomrule
  \end{tabular}
\end{table}

Across these controls, the initial strong advantage observed under checkpoint initialization does not persist. Loss differences collapse under fair initialization, and activity differences disappear under degree-preserving rewiring. The remaining runtime differences are modest in magnitude.
\section{Control Analysis}

\subsection{Limitations of the Original Comparison}

The original comparison combines two assumptions that favor the connectome model:
\begin{enumerate}[leftmargin=1.5em]
  \item both graphs are initialized from a checkpoint trained on the connectome topology, and
  \item the control graph matches only global counts and self-loops, without preserving directed degree sequence.
\end{enumerate}
Each assumption introduces a distinct source of bias. Checkpoint initialization transfers parameters adapted to the connectome topology, while a naive random null alters degree structure that can independently affect activity distribution and runtime \citep{newman2001randomgraphs,maslov2002specificity,milo2002network}. These factors prevent the original comparison from isolating topology alone.

\subsection{Effect of Progressive Controls}

\begin{figure}[t]
  \centering
  \includegraphics[width=\linewidth]{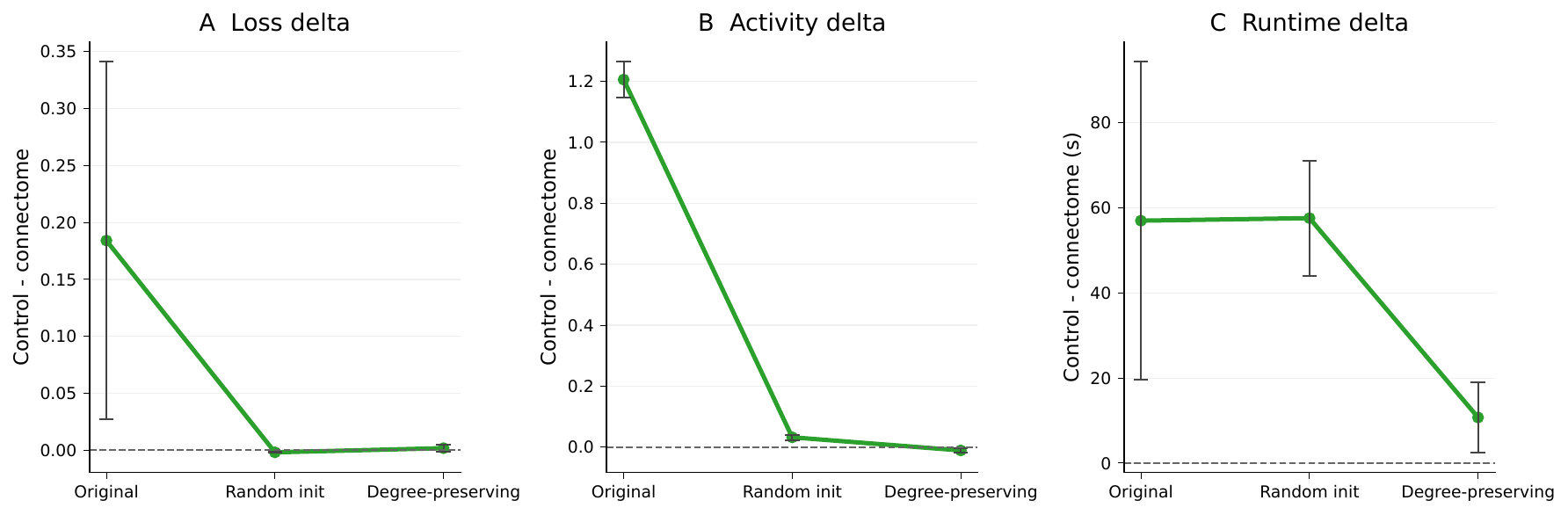}
  \caption{Control progression. Each bar shows control-minus-connectome differences across successive control levels. Loss differences vanish under shared random initialization, and activity differences vanish under the degree-preserving null.}
  \label{fig:revision-controls}
\end{figure}

Figure~\ref{fig:revision-controls} summarizes the progression of control-minus-connectome differences across the study. Three transitions are decisive:
\begin{enumerate}[leftmargin=1.5em]
  \item Under checkpoint initialization, all primary metrics favor the connectome.
  \item Under shared random initialization, the loss difference collapses to approximately zero.
  \item Under a degree-preserving null, the activity difference also disappears and slightly reverses.
\end{enumerate}

Thus, the two dominant effects in the original comparison—lower loss and lower activity—do not persist under stricter controls. The remaining runtime difference is smaller and does not follow the same pattern.

\subsection{Role of Degree-Preserving Null Models}

Matching only global graph statistics is insufficient for isolating topology effects. Directed degree sequence influences routing patterns, activity concentration, and memory access structure. Degree-preserving rewiring therefore provides a substantially stronger null by retaining this structural constraint while randomizing higher-order connectivity. Although it does not preserve all graph properties, it removes a major source of mismatch present in naive random controls.

\subsection{Ensemble Robustness}

The degree-preserving result is stable across multiple independent rewired graphs. Across five samples and three seeds per sample, the connectome does not recover the original advantage. Loss differences remain near zero, activity differences are slightly negative, and runtime differences remain modest. This consistency indicates that the corrected outcome is not driven by a single rewiring instance.

\subsection{Initialization and Dynamical Scale}

To test whether residual differences arise from trivial scale mismatches, we compare pre-training activity and gradient norms under the same initialization procedure. Connectome and degree-preserving graphs begin at closely matched scales, with mean absolute activity differing by only $1.64\%$ and similar gradient magnitudes. This rules out large initial-scale differences as an explanation for the corrected results.

\subsection{Interpretation of Runtime Differences}

Elapsed time remains systematically lower for the connectome, but its interpretation is limited. While graph size, decoder, and optimizer are matched, runtime in this implementation depends on adjacency ordering, memory locality, and parameter-sharing access patterns. Activity magnitude may also affect numerical behavior indirectly. Accordingly, runtime is reported as an empirical observation but is not treated as evidence of a topology-specific computational advantage.

\subsection{Resulting Claim}

Under progressively stricter controls, the original topology advantage does not persist. The supported conclusion is therefore:

\begin{quote}
Apparent topology advantages in connectome-constrained neural networks are highly sensitive to initialization and null-model design, and do not robustly persist under degree-preserving controls.
\end{quote}
\section{Mechanism Analysis}

This section characterizes activity patterns observed in the connectome-versus-naive-random comparison. The analysis is descriptive and is restricted to that weak-control setting; it is not used to infer a causal topology advantage under the corrected controls.

Three consistent patterns are observed. First, total activity is substantially lower in the connectome model than in the naive random model. Second, node activity and edge-usage proxies are less concentrated in the connectome: node-activity Gini is $0.432$ versus $0.469$, and edge-usage Gini is $0.779$ versus $0.823$. Third, temporal summaries are mixed: the connectome exhibits lower total activity and lower node-wise variance, but does not consistently show smoother timestep-to-timestep dynamics.

\begin{table}[t]
  \centering
  \caption{Descriptive activity and usage metrics for the connectome and naive random graphs under the weak-control comparison. These values summarize behavior in that setting and are not interpreted as causal evidence under the corrected controls.}
  \label{tab:mechanism}
  \begin{tabular}{lrr}
    \toprule
    Metric & Connectome & Naive random \\
    \midrule
    Mean absolute activity & 0.6648 $\pm$ 0.0481 & 1.7355 $\pm$ 0.0150 \\
    Node activity Gini & 0.4319 $\pm$ 0.0372 & 0.4694 $\pm$ 0.0033 \\
    Top 10\% node activity fraction & 0.2807 $\pm$ 0.0308 & 0.3333 $\pm$ 0.0027 \\
    Edge usage Gini & 0.7786 $\pm$ 0.0080 & 0.8232 $\pm$ 0.0025 \\
    Top 10\% edge usage fraction & 0.6031 $\pm$ 0.0159 & 0.6884 $\pm$ 0.0034 \\
    Mean total activity over time & 30{,}358.9 $\pm$ 2{,}195.8 & 79{,}258.3 $\pm$ 686.7 \\
    \bottomrule
  \end{tabular}
\end{table}

Taken together, these results indicate that, under the weak-control comparison, the connectome distributes activity more evenly across nodes and edges while operating at a lower overall activity scale. This pattern is consistent with a more distributed routing regime relative to the naive random graph. However, these observations do not establish that connectome topology confers a causal advantage under the corrected control conditions. Under degree-preserving controls, the corresponding performance differences do not persist.
\section{Discussion}

This study isolates how conclusions about topology change under progressively stricter controls. The central observation is that an initially strong connectome advantage does not persist once checkpoint initialization and weak null models are removed. Two factors account for this shift. First, initialization. When both graphs are recovered from a checkpoint trained on the connectome topology, the comparison incorporates parameters already adapted to that structure. Under a shared from-scratch initialization, the loss advantage disappears. Second, null-model design. A self-loop-matched random graph does not preserve directed degree sequence and therefore alters broad structural properties that influence activity and routing. When the control preserves degree sequence, the activity advantage also disappears. These findings highlight a general issue in sparse-network comparisons. Differences attributed to ``topology'' can arise from multiple structural and procedural factors, including degree sequence, motif distribution, self-loops, and initialization history \citep{newman2001randomgraphs,maslov2002specificity,rubinov2010complex}. Treating null-model construction as an explicit part of the experimental design is therefore essential for isolating topology effects.

The remaining runtime difference is reproducible but limited in interpretability. Although graph size and parameter count are matched, execution time depends on implementation details such as adjacency ordering, memory locality, and parameter-sharing access patterns. These factors are not fully disentangled in the present setup, so runtime is reported as an empirical observation rather than as evidence of a topology-specific computational advantage. More broadly, the contribution is methodological. Connectome-constrained models provide a controlled setting in which to test how structural assumptions influence learning, but only when initialization and null-model choices are made explicit and systematically varied. Under such controls, the present results indicate that apparent topology advantages are not robust. This shifts the emphasis from identifying favorable structures to designing comparisons that cleanly isolate their effects.
\section{Limitations}

The conclusions of this study are intentionally narrow and apply within a specific experimental regime.

\begin{enumerate}[leftmargin=1.5em]

  \item \textbf{Single task family.} All results are obtained on the MovingEdge direction-decoding task within the flyvis Stage 3 setup. The extent to which the same behavior appears on other tasks or domains remains open.

  \item \textbf{Short training horizon.} The matched-step protocol evaluates models after 5 and 10 optimization steps. This isolates early-learning dynamics but does not address longer training trajectories or final convergence.

  \item \textbf{Limited seeds.} Core comparisons use three optimization seeds. Robustness is primarily assessed through multiple degree-preserving graph samples rather than a large number of independent training runs.

  \item \textbf{Implementation-bound parameter sharing.} Rewiring modifies adjacency while preserving the flyvis metadata tables that define parameter sharing. As a result, the comparison isolates topology within a fixed implementation but does not fully separate topology from all parameter-sharing effects.

  \item \textbf{Partial null-model control.} The degree-preserving null enforces exact matching of directed in-degree, out-degree, edge count, and self-loops, but does not match higher-order structure such as clustering, assortativity, motif statistics, or spatial organization. The rewiring budget also leaves a fraction of edges unchanged, so the null remains partially correlated with the original graph.

  \item \textbf{Uncontrolled spectral properties.} Initial activity and gradient scales are comparable across graph conditions, but spectral radius and related operator-level properties are not explicitly matched.

  \item \textbf{Inconclusive structured nulls.} Small-world and ring-lattice controls became non-finite before a controlled comparison could be completed and therefore do not provide informative counterexamples.

  \item \textbf{Descriptive mechanism analysis.} Mechanism metrics characterize activity patterns under weak-control conditions but are not used to infer causal effects under the corrected controls.

\end{enumerate}

These constraints define the scope of the results. The conclusions should therefore be interpreted as a controlled statement about early-learning behavior under matched sparsity and parameterization, rather than as a general claim about topology across architectures or tasks.
\section{Conclusion}

This study re-examines an apparent connectome advantage under progressively stricter controls. While initial comparisons suggested that a connectome-constrained network outperformed a matched random graph in early optimization, activity, and runtime, these differences do not persist once checkpoint initialization and weak null models are removed. Under shared random initialization and degree-preserving controls, both loss and activity differences collapse to near zero.

The resulting conclusion is methodological. Apparent topology advantages in connectome-constrained neural networks depend critically on initialization and null-model design, and are not robust under degree-preserving controls. 

More broadly, these results highlight the importance of treating null-model construction as an explicit component of experimental design. In sparse network comparisons, conclusions about topology can change substantially once appropriate controls are applied. Connectome-constrained models therefore remain useful as controlled testbeds for investigating structural effects, provided that initialization and null-model choices are carefully specified.
\section{Repository Structure and Access}

The source code, data, and results of this study are publicly available in the following GitHub repository:

\href{https://github.com/nalin-dhiman/Connectome-Constrained-Neural-Networks}{Github}

This repository serves as both the technical foundation for reproducing the experiments and the transparent storage for all data and results related to this work.

\bibliographystyle{plainnat}
\bibliography{references}

\end{document}